\documentclass[copyright,creativecommons]{eptcs}

\providecommand{\event}{ACL2-2020}

\usepackage{alltt}
\usepackage{amssymb}
\usepackage{enumitem}
\usepackage[T1]{fontenc}
\usepackage{hyperref}
\usepackage{inconsolata}
\usepackage{tikz}
\usepackage{underscore}
\usepackage{url}

\newcommand{\setFT}[2]{\{#1,\ldots,#2\}} 
\newcommand{\setsum}{\uplus} 
\newcommand{\Seq}[1]{#1^\star} 

\newcommand{\rlpstring}[1]{[#1]} 
\newcommand{\rlpstringO}{\rlpstring{\,}} 
\newcommand{\rlplist}[1]{\langle#1\rangle} 
\newcommand{\rlplistO}{\rlplist{\,}} 
\newcommand{\Rlptree}{\mathbb{T}} 
\newcommand{\Rlplist}{\mathbb{L}} 
\newcommand{\Rlpstring}{\mathbb{B}} 
\newcommand{\Byte}{\mathbb{O}} 
\newcommand{\rlpencbytes}{R_\mathrm{b}} 
\newcommand{\rlpenclist}{R_\mathrm{l}} 
\newcommand{\rlpenctree}{\mathtt{RLP}} 
\newcommand{\rlpenctrees}{s} 
\newcommand{\bendian}{\mathtt{BE}} 
\newcommand{\undefined}{\varnothing} 

\newcommand{\len}{l} 
\newcommand{\lenlen}{\mathit{ll}} 

\newcommand{\pyth}[1]{\texttt{#1}} 

\newcommand{\acl}[1]{\texttt{#1}} 
\newenvironment{bacl}{\small\begin{alltt}}{\end{alltt}} 

\newcommand{\secref}[1]{Section \ref{#1}}
\newcommand{\figref}[1]{Figure \ref{#1}}

\newcommand{\citecode}[2]{\cite[Path \href{#1}{\texttt{#2}}]{acl2-code}}
\newcommand{\citeman}[2]{\cite[Topic \href{#1}{\texttt{#2}}]{acl2-manual}}
\newcommand{\citewiki}[2]{\cite[Page \href{#1}{`#2'}]{eth-wiki}}

\begin{document}


\title{Ethereum's Recursive Length Prefix in ACL2}

\author{Alessandro Coglio
        \institute{Kestrel Institute \\ \url{http://www.kestrel.edu}}}

\def\titlerunning{Ethereum RLP in ACL2}
\def\authorrunning{Alessandro Coglio}

\maketitle

\begin{abstract}
Recursive Length Prefix (RLP)
is used to encode a wide variety of data in Ethereum,
including transactions.
The work described in this paper provides
a formal specification of RLP encoding and
a verified implementation of RLP decoding,
developed in the ACL2 theorem prover.
This work has led to
improvements to the Ethereum documentation
and additions to the Ethereum test suite.
\end{abstract}


\section{Problem and Contribution}
\label{sec:prob-contrib}

Errors in cryptocurrency code may lead to particularly direct financial losses.
This applies not only to smart contracts,
but also to the underlying execution engines,
to wallets,
and to other critical components.

In Ethereum \cite{eth-www},
Recursive Length Prefix (RLP)
\citewiki{https://github.com/ethereum/wiki/wiki/RLP}{RLP}
\cite[Appendix B]{eth-yp}
is used to encode a wide variety of data,
including transactions.
It is thus important for this fundamental building block
to be specified precisely and implemented correctly.

The work described in this paper contributes to this goal by providing
\emph{a formal specification of RLP encoding and
a verified implementation of RLP decoding},
developed in the ACL2 theorem prover \cite{acl2-www}.
The development is available
\citecode{https://github.com/acl2/acl2/tree/master/books/kestrel/ethereum/rlp}{books/kestrel/ethereum/rlp}
and thoroughly documented
\citeman{http://www.cs.utexas.edu/users/moore/acl2/manuals/current/manual/?topic=ETHEREUM\_\_\_\_RLP}{rlp}.
Some excerpts of the development shown in this paper
are slightly simplified for brevity.

This work has led to improvements
to the Ethereum Yellow Paper \cite{eth-yp}
and to the Ethereum Wiki \cite{eth-wiki},
which are major components of the Ethereum documentation.
It has also led to additions to the Ethereum test suite \cite{eth-tests},
which contains tests for all Ethereum implementations.
See \secref{sec:related} for details.

This work is part of an ongoing effort to develop, in ACL2,
a formal specification and a verified implementation
of a complete Ethereum client \cite{acl2-eth}.
This formal specification will also be useful for formally verifying
existing client implementations,
smart contracts at the level of the EVM (Ethereum Virtual Machine),
and the compilation of higher-level programming languages to EVM code.


\section{Background}
\label{sec:back}

Ethereum uses RLP to encode transactions,
which may include smart contract code to run on the EVM.
It also uses RLP to encode blocks,
whose hashing plays a critical role in the blockchain paradigm.
Thus, errors in RLP encoders and decoders may lead, among other problems,
to unexpected smart contract code,
or to incorrect block verification or mining.

RLP is specified informally in the Ethereum wiki (`WK' for short)
\citewiki{https://github.com/ethereum/wiki/wiki/RLP}{RLP}
and more formally in the Ethereum Yellow Paper (`YP' for short)
\cite[Appendix B]{eth-yp}.

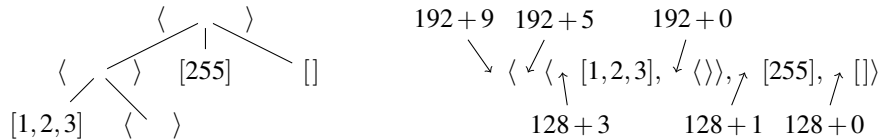
\begin{figure}
\centering
\begin{tikzpicture}
\node (branch0) at (  0  ,   0  ) {\small $\rlplist{\hspace{30pt}}$};
\node (branch1) at (-40pt, -20pt) {\small $\rlplist{\hspace{25pt}}$};
\node (branch2) at (-20pt, -40pt) {\small $\rlplist{\hspace{15pt}}$};
\node (leaf0)   at (-60pt, -40pt) {\small $\rlpstring{1,2,3}$};
\node (leaf1)   at (  0  , -20pt) {\small $\rlpstring{255}$};
\node (leaf2)   at ( 40pt, -20pt) {\small $\rlpstringO$};
\coordinate (branch0) at (  0  ,   0  );
\coordinate (branch1) at (-40pt, -20pt);
\coordinate (branch2) at (-20pt, -40pt);
\draw (branch0) -- (branch1);
\draw (branch0) -- (leaf1);
\draw (branch0) -- (leaf2);
\draw (branch1) -- (leaf0);
\draw (branch1) -- (branch2);
\filldraw[fill=white, draw=white] (branch0) circle (3pt);
\filldraw[fill=white, draw=white] (branch1) circle (3pt);
\filldraw[fill=white, draw=white] (branch2) circle (3pt);
\node at (100pt, -20pt) [anchor=west]
 {\small
  $\quad\rlplist{\quad\rlplist{\quad\rlpstring{1,2,3},
                               \quad\rlplistO},
                      \quad\rlpstring{255},
                      \quad\rlpstringO}$};
\coordinate (prefix-branch0) at (108pt, -20pt);
\coordinate (prefix-branch1) at (122pt, -20pt);
\coordinate (prefix-branch2) at (178pt, -20pt);
\coordinate (prefix-leaf0)   at (135pt, -20pt);
\coordinate (prefix-leaf1)   at (204pt, -20pt);
\coordinate (prefix-leaf2)   at (240pt, -20pt);
\node (len-branch0) at ( 94pt,   0pt) {\small $192+9$};
\node (len-branch1) at (132pt,   0pt) {\small $192+5$};
\node (len-branch2) at (184pt,   0pt) {\small $192+0$};
\node (len-leaf0)   at (138pt, -40pt) {\small $128+3$};
\node (len-leaf1)   at (196pt, -40pt) {\small $128+1$};
\node (len-leaf2)   at (234pt, -40pt) {\small $128+0$};
\draw [->] (len-branch0) -- (prefix-branch0);
\draw [->] (len-branch1) -- (prefix-branch1);
\draw [->] (len-branch2) -- (prefix-branch2);
\draw [->] (len-leaf0) -- (prefix-leaf0);
\draw [->] (len-leaf1) -- (prefix-leaf1);
\draw [->] (len-leaf2) -- (prefix-leaf2);
\end{tikzpicture}
\caption{\label{fig:rlp-tree-example}
An example RLP tree and its encoding.}
\end{figure}

RLP encodes nested sequences of bytes into flat sequences of bytes
that can be decoded back into the original nested sequences.
These nested sequences are finitely branching ordered trees
with flat sequences of bytes at their leaf nodes
and no additional information at the branching nodes.
An example is
$\rlplist{\rlplist{\rlpstring{1,2,3},
                   \rlplistO},
          \rlpstring{255},
          \rlpstringO}$,
where $\rlpstring{\ldots}$ denote leaf nodes
and $\rlplist{\ldots}$ denote branching nodes;
this tree is depicted in \figref{fig:rlp-tree-example} (left).
The leaf tree $\rlpstringO$ with the empty sequence of bytes
differs from the branching tree $\rlplistO$ with no subtrees.%
\footnote{In this paper,
a `leaf tree' is one that consists of just one leaf node
(which is also necessarily the root),
while a `branching tree' is one that has (at least)
a branching node as root.}

WK calls the trees `items',
the branching trees `lists',
and the leaf trees `strings' or `byte arrays';
it uses Python notations
\pyth{"..."} for strings
(with the implicit assumption that characters consist of 8 bits)
and \pyth{[...,...]} for lists;
it also uses lists \pyth{[...,...]} of numbers and characters
for flat sequences of bytes.
YP uses a more explicit tree terminology;
it uses a mathematical notation $(\ldots,\dots)$
for both leaf and branching nodes, as well as for flat sequences of bytes.
This paper uses the notations $\rlpstring{\ldots}$ and $\rlplist{\ldots}$
shown earlier and in \figref{fig:rlp-tree-example}.

RLP prescribes to encode non-negative integers
by first representing them as
base-256 big-endian byte arrays without leading zeros
(not even a 0 byte for the integer 0,
which is represented as the empty byte array),
and then encoding those as any other leaf trees.
Other than that, RLP does not prescribe encodings for any data types,
delegating that to the ``users'' of RLP.

Leaf and branching trees are encoded into byte arrays
by recursively adding a few extra bytes before most nodes
to indicate the kind of node (leaf or branching)
and the length of the subsequent bytes.
These extra bytes tell the decoder how to reconstruct the nodes of the tree.

Leaf trees $\rlpstring{\ldots}$ are encoded
by encoding their byte arrays as follows.
Singleton byte arrays whose only byte is below 128
are encoded as themselves, without extra bytes.
Other byte arrays whose length is in the range 0--55
have an extra starting byte in the range 128--183,
which is 128 plus the length.
Longer byte arrays have from 1 to 8 extra bytes
that contain the base-256 big-endian no-leading-zeros length,
preceded by an extra byte in the range 184--191
that indicates the number of the base-256 big-endian bytes---%
184 for 1, 185 for 2, \ldots, and 191 for 8.

Branching trees $\rlplist{\ldots}$ are encoded
by first recursively encoding the subtrees
and concatenating all the encodings,
and then adding one or more extra bytes to indicate
the length of the concatenated encodings:
if the length is in the range 0--55,
there is a single extra byte in the range 192--247,
which is 192 plus the length;
otherwise, there are from 1 to 8 extra bytes
that contain the base-256 big-endian no-leading-zero length,
preceded by an extra byte in the range 248--255
that indicates the number of the base-256 big-endian bytes---%
248 for 1, 249 for 2, \ldots, and 255 for 8.

For instance,
$\rlplist{\rlplist{\rlpstring{1,2,3},
                   \rlplistO},
          \rlpstring{255},
          \rlpstringO}$
is encoded as
$\rlpstring{201,197,131,1,2,3,192,129,255,128}$.
This is illustrated in \figref{fig:rlp-tree-example} (right).

There is a symmetry between leaf and branch encodings
with respect to small and large lengths:
the first 56 values of the 64 values in the range 128--191 or 192--255
are used for small lengths from 0 to 55,
while the remaining 8 values are used for large lengths
up to $2^{64}-1$, i.e.\ the maximum value representable in
8 digits in base 256, since $256^8=2^{64}$.
However, leaf trees have an additional shorter encoding,
when they consist of single bytes below 128.


\section{RLP Encoding}


\subsection{Tree Structures}

As explained in \secref{sec:back},
RLP encodes trees (i.e.\ nested byte sequences)
into flat byte sequences (i.e.\ byte arrays).
These trees are formalized in \figref{fig:rlp-trees-acl2},
using the FTY library
\citeman{http://www.cs.utexas.edu/users/moore/acl2/manuals/current/manual/?topic=ACL2____FTY}{fty}.

\begin{figure}
\centering
\begin{minipage}{0.55\textwidth}
\begin{bacl}
(fty::deftypes rlp-trees
  (fty::deftagsum rlp-tree
    (:leaf ((bytes byte-list)))
    (:branch ((subtrees rlp-tree-list))))
  (fty::deflist rlp-tree-list :elt-type rlp-tree))
\end{bacl}
\end{minipage}
\caption{\label{fig:rlp-trees-acl2}
Formalization of RLP trees in ACL2.}
\end{figure}

The \acl{fty::deftagsum} defines
a tagged sum type (disjoint union) called \acl{rlp-tree}.
Leaf trees are tagged by \acl{:leaf};
branching trees are tagged by \acl{:branch}.
A leaf tree has a single component called \acl{bytes}
of type \acl{byte-list},
which consists of lists of natural numbers below 256
(its definition is not shown).
A branching tree has a single component called \acl{subtrees}
of type \acl{rlp-tree-list},
defined by the \acl{fty::deflist} as consisting of lists of trees;
\acl{:elt-type} specifies the element type of the lists.
The surrounding \acl{fty::deftypes}
introduces \acl{rlp-tree} and \acl{rlp-tree-list}
as mutually recursive types,
with \acl{rlp-trees} as the name of the ensemble.

In more detail,
the \acl{fty::deftagsum} introduces
a recognizer \acl{rlp-treep},
a fixer \acl{rlp-tree-fix},
constructors \acl{rlp-tree-leaf} and \acl{rlp-tree-branch}
for leaf and branching trees,
accessors \acl{rlp-tree\-leaf->bytes}
and \acl{rlp-tree-branch->subtrees}
for leaf and branching trees,
and several theorems about these functions,
which are all guard-verified.
The \acl{fty::deflist} introduces
a recognizer \acl{rlp\-tree-listp},
a fixer \acl{rlp-tree-list-fix},
and theorems about these functions and existing list functions;
no specific constructors or accessors are introduced by \acl{fty::deflist},
since the generic ones for lists can be used.

In essence, \figref{fig:rlp-trees-acl2} defines the set $\Rlptree$ of trees
as a least fixpoint of the recursive set equation
$\Rlptree = \Seq{\setFT{0}{255}} \setsum \Seq{\Rlptree}$,
where $\setFT{0}{255}$ is the set of bytes,
$\Seq{X}$ is the set of all the finite sequences of elements in $X$
(Kleene star),
and $X \setsum Y$ is the disjoint union of $X$ and $Y$.
This is consistent with the set-theoretic notation in YP,
which uses
$\Byte$ for $\setFT{0}{255}$,
$\Rlpstring$ for $\Seq{\Byte}$, and
$\Rlplist$ for $\Seq{\Rlptree}$.


\subsection{Encoding Functions}
\label{sec:encoders}

The encoding of RLP trees into byte arrays
is formalized in \figref{fig:rlp-encode-acl2}.

The function \acl{rlp-encode-bytes}
formalizes the encoding of the byte arrays at the leaves of RLP trees,
using the \acl{define} enhancement of \acl{defun}
\citeman{http://www.cs.utexas.edu/users/moore/acl2/manuals/current/manual/?topic=ACL2____DEFINE}{define},
which supports type annotations
for arguments (abbreviating guards)
and results (abbreviating theorems).
The function \acl{rlp-encode-bytes}
takes as input a list of bytes
and returns as output a pair
\citeman{http://www.cs.utexas.edu/users/moore/acl2/manuals/current/manual/?topic=ACL2____MV}{mv}
consisting of a boolean error flag
and a list of bytes that is the RLP encoding of the argument.
The error flag is \acl{t} exactly when
the input consists of $2^{64}$ or more bytes:
in this case, the input cannot be RLP-encoded,
and the second component of the output is just \acl{nil}, but irrelevant.
Otherwise, the error flag is \acl{nil}
and the second component of the output is the encoding.

First, \acl{rlp-encode-bytes} fixes
the argument to be a list of bytes via the fixer \acl{byte-list-fix}
(a no-op under the guard),
using the \acl{b*} enhancement of \acl{let*}
\citeman{http://www.cs.utexas.edu/users/moore/acl2/manuals/current/manual/?topic=ACL2____B_A2}{b*}.
Then there are four cases:
\begin{enumerate}[nosep]
\item
If the list consists of one byte
and that byte is below 128,
the operation is successful (i.e.\ the \acl{error?} result is \acl{nil})
and the encoding is the singleton list of the byte itself.
\item
Otherwise, if the list consists of $\len < 56$ bytes,
the operation is successful
and the encoding is obtained by prepending the byte $128 + \len$
to the list of bytes.
\item
Otherwise, if the list consists of $\len < 2^{64}$ bytes,
the operation is successful
and the encoding is obtained by prepending
(i) the byte $183 + \lenlen$
to the concatenation of
(ii) the base-256 big-endian no-leading-zeros representation of $\len$
of length $\lenlen$ (`length of length') and
(iii) the initial list of bytes.
The library function \acl{nat=>bebytes*} turns a natural number
into a list of bytes as big-endian digits in base 256,
without leading zeros.
\item
Otherwise, the operation fails:
the list of bytes is too long to be encoded.
To encode a list of $2^{64}$ or more bytes,
9 or more base-256 digits would be needed,
i.e.\ it would be $\lenlen \geq 9$,
and therefore the first byte would be 192 or more,
overlapping with the encoding of branching trees
and preventing a decoder from discriminating leaf and branching trees
by the first byte of an encoding.
\end{enumerate}

\begin{figure}
\centering
\begin{minipage}{0.86\textwidth}
\begin{bacl}
(define rlp-encode-bytes ((bytes byte-listp))
  :returns (mv (error? booleanp) (encoding byte-listp))
  (b* ((bytes (byte-list-fix bytes)))
    (cond ((and (= (len bytes) 1) (< (car bytes) 128)) (mv nil bytes))
          ((< (len bytes) 56) (mv nil (cons (+ 128 (len bytes)) bytes)))
          ((< (len bytes) (expt 2 64))
           (b* ((be (nat=>bebytes* (len bytes))))
             (mv nil (cons (+ 183 (len be)) (append be bytes)))))
          (t (mv t nil)))))

(defines rlp-encode-trees
  (define rlp-encode-tree ((tree rlp-treep))
    :returns (mv (error? booleanp) (encoding byte-listp))
    (rlp-tree-case tree
     :leaf (rlp-encode-bytes tree.bytes)
     :branch (b* (((mv error? encoding) (rlp-encode-tree-list tree.subtrees))
                  ((when error?) (mv t nil)))
               (cond ((< (len encoding) 56)
                      (mv nil (cons (+ 192 (len encoding)) encoding)))
                     ((< (len encoding) (expt 2 64))
                      (b* ((be (nat=>bebytes* (len encoding))))
                        (mv nil (cons (+ 247 (len be)) (append be encoding)))))
                     (t (mv t nil))))))
  (define rlp-encode-tree-list ((trees rlp-tree-listp))
    :returns (mv (error? booleanp) (encoding byte-listp))
    (b* (((when (endp trees)) (mv nil nil))
         ((mv error? encoding1) (rlp-encode-tree (car trees)))
         ((when error?) (mv t nil))
         ((mv error? encoding2) (rlp-encode-tree-list (cdr trees)))
         ((when error?) (mv t nil)))
      (mv nil (append encoding1 encoding2)))))
\end{bacl}
\end{minipage}
\caption{\label{fig:rlp-encode-acl2}
Formalization of RLP encoding in ACL2.}
\end{figure}

The function \acl{rlp-encode-bytes} formalizes
the function $\rlpencbytes$ in YP,
which returns one result:
either the encoding (a byte sequence),
or $\undefined$ if the input byte sequence cannot be encoded.
In the ACL2 formulation,
it is more convenient to return two results,
so that each result always has the same type.
The function \acl{nat=>bebytes*}, mentioned above,
formalizes the function $\bendian$ in YP.

The function \acl{rlp-encode-tree}
formalizes the encoding of (both branching and leaf) trees.
It takes as input a tree
and returns as output a pair consisting of
a boolean error flag
and a list of bytes that is the RLP encoding of the argument---%
the same output types as \acl{rlp-encode-bytes}.

First, \acl{rlp-encode-tree} performs a case analysis on the argument
via the macro \acl{rlp-tree-case},
generated by the \acl{fty::deftagsum} in \figref{fig:rlp-trees-acl2}.
If \acl{tree} is a leaf tree,
\acl{rlp-encode-bytes} is called on the list of bytes in the leaf;
the variable \acl{tree.bytes} is bound
to \acl{(rlp-tree-leaf->bytes tree)}
by \acl{rlp-tree-case}.
If instead \acl{tree} is a branching tree,
\acl{rlp-encode-tree-list} is called to encode each subtree
and concatenate their encodings (more details below);
the variable \acl{tree.subtrees} is bound
to \acl{(rlp-tree-branch->subtrees tree)}
by \acl{rlp-tree-case}.
The two results of \acl{rlp-encode-tree-list}
are simultaneously bound to \acl{error?} and \acl{encoding}
via \acl{b*}'s support for binding patterns such as \acl{(mv ...)}.
If \acl{rlp-encode-tree-list} returns an error,
\acl{rlp-encode-tree} returns an error too,
via \acl{b*}'s early-exit construct \acl{((when ...) ...)}:
if any subtree cannot be encoded, the tree cannot be encoded either.
Otherwise, there are three cases:
\begin{enumerate}[nosep]
\item
If the concatenated subtree encodings consist of $\len < 56$ bytes,
the overall operation is successful
and the overall encoding is obtained by prepending the byte $192 + \len$
to the concatenated subtree encodings.
\item
Otherwise,
if the concatenated subtree encodings consist of $\len < 2^{64}$ bytes,
the overall operation is successful
and the overall encoding is obtained by prepending
(i) the byte $247 + \lenlen$
to the concatenation of
(ii) the base-256 big-endian no-leading-zeros representation of $\len$
of length $\lenlen$ and
(iii) the concatenated subtree encodings.
\item
Otherwise, the overall operation fails,
because the tree is too large to be encoded.
To encode a tree whose concatenated subtree encodings
consist of $2^{64}$ or more bytes,
9 or more base-256 digits would be needed,
i.e.\ it would be $\lenlen \geq 9$,
and therefore the first byte would be 256 or more,
which would not actually be a byte.
\end{enumerate}

The function \acl{rlp-encode-tree-list}
formalizes the encoding of a list of (sub)trees
and the concatenation of the resulting encodings.
It takes as input a list of trees
and returns as output a pair consisting of
a boolean error flag
and a list of bytes that are the concatenated encodings of the argument trees;
these are the same output types as \acl{rlp-encode-tree}.
When the list of trees is empty,
the result is the empty list of bytes \acl{nil}.
Otherwise, the first tree is encoded,
the remaining list of trees is encoded,
and the two resulting lists of bytes are concatenated.
If any tree in the list cannot be encoded, an error is returned.

The functions \acl{rlp-encode-tree} and \acl{rlp-encode-tree-list}
are mutually recursive.
The macro \acl{defines}
\citeman{http://www.cs.utexas.edu/users/moore/acl2/manuals/current/manual/?topic=ACL2____DEFINES}{defines}
groups mutually recursive functions
that are introduced via \acl{define}
(and also provides some enhancements over \acl{mutual-recursion}),
with \acl{rlp-encode-trees} as the name of the ensemble.
Termination is proved automatically,
based on the decreasing size of the argument trees,
which is explicitly supplied as measure (not shown).

The function \acl{rlp-encode-tree}
formalizes the function $\rlpenctree$ in YP;
the \acl{:branch} case of the definition of \acl{rlp-encode-tree}
formalizes the function $\rlpenclist$ in YP.
The function \acl{rlp-encode-tree-list}
formalizes the function $\rlpenctrees$ in YP.

The functions \acl{rlp-encode-...} are guard-verified;
their guard verification proofs are essentially automatic,
with just a hint (not shown)
to locally enable a rewrite rule
that may be somewhat expensive to be always enabled.
The proofs of the result type theorems of the \acl{rlp-encode-...} functions
are also proved essentially automatically,
with just a hint (not shown)
to locally enable a definition that is normally kept disabled
and to locally enable a linear arithmetic rule
that may be somewhat expensive to be always enabled.
All these proofs make use of existing library rules
about the functions called by the \acl{rlp-encode-...} functions.

The \acl{rlp-encode-...} functions
provide a high-level specification of RLP encoding
that also happens to be executable.
The definitions of these functions correspond very closely
to the definitions in YP.
These definitions are also similar to the Python reference code in WK.


\subsection{Valid Encodings}
\label{sec:encodings}

Valid encodings are formalized in \figref{fig:valid-rlp-encodings-acl2},
using the \acl{define-sk} enhancement
\citeman{http://www.cs.utexas.edu/users/moore/acl2/manuals/current/manual/?topic=STD____DEFINE-SK}{define-sk}
of \acl{defun-sk}
\citeman{http://www.cs.utexas.edu/users/moore/acl2/manuals/current/manual/?topic=ACL2____DEFUN-SK}{defun-sk},
which provides conveniences similar to \acl{define}.

\begin{figure}
\centering
\begin{minipage}{0.83\textwidth}
\begin{bacl}
(define-sk rlp-tree-encoding-p ((encoding byte-listp))
  :returns (yes/no booleanp)
  (exists (tree) (and (rlp-treep tree)
                      (b* (((mv error? encoding1) (rlp-encode-tree tree)))
                        (and (not error?)
                             (equal encoding1 (byte-list-fix encoding))))))
  :skolem-name rlp-tree-encoding-witness)
\end{bacl}
\end{minipage}
\caption{\label{fig:valid-rlp-encodings-acl2}
Formalization of valid RLP encodings in ACL2.}
\end{figure}

The predicate \acl{rlp-tree-encoding-p} returns \acl{t}
exactly on the byte arrays that encode some trees.
Roughly speaking, this predicate characterizes
the image of \acl{rlp-encode-tree},
restricted to the second result of that function,
and subject to the constraint that the first result is \acl{nil}.
This predicate has a declarative, non-executable definition.


\subsection{Decodability Properties}
\label{sec:decodability}

RLP encodings are decodable,
i.e.\ trees can be recovered from their encodings---%
a basic requirement for any encoding method.
This is expressed by the theorems in
\figref{fig:rlp-decodable-acl2}:
(i) the RLP encoding function is injective,
i.e.\ no two distinct trees have the same encoding; and
(ii) the RLP encoding function is prefix-unambiguous,
i.e.\ no valid tree encoding is a strict prefix of another one.
The second property ensures the ability to decode
from byte streams without ``end-of-encoding'' markers:
if a valid encoding could be a strict prefix of another valid encoding,
then after reading the former,
a decoder could either stop there
or proceed to decode a longer encoding,
giving rise to an ambiguity.

\begin{figure}
\centering
\begin{minipage}{0.68\textwidth}
\begin{bacl}
(defthm rlp-encode-tree-injective
  (implies (and (not (mv-nth 0 (rlp-encode-tree x)))
                (not (mv-nth 0 (rlp-encode-tree y))))
           (equal (equal (mv-nth 1 (rlp-encode-tree x))
                         (mv-nth 1 (rlp-encode-tree y)))
                  (equal (rlp-tree-fix x) (rlp-tree-fix y)))))

(defthm rlp-encode-tree-unamb-prefix
  (implies (and (not (mv-nth 0 (rlp-encode-tree x)))
                (not (mv-nth 0 (rlp-encode-tree y))))
           (equal (prefixp (mv-nth 1 (rlp-encode-tree x))
                           (mv-nth 1 (rlp-encode-tree y)))
                  (equal (mv-nth 1 (rlp-encode-tree x))
                         (mv-nth 1 (rlp-encode-tree y))))))
\end{bacl}
\end{minipage}
\caption{\label{fig:rlp-decodable-acl2}
ACL2 theorems asserting the decodability of RLP encodings.}
\end{figure}

Typically, the injectivity of a function $f$ is stated as
$[x \neq y \Longrightarrow f(x) \neq f(y)]$,
with $x$ and $y$ universally quantified,
or equivalently as
$[f(x) = f(y) \Longrightarrow x = y]$.
Because $[x = y \Longrightarrow f(x) = f(y)]$ is always trivially true,
injectivity can be equivalently stated as
$[f(x) = f(y) \Longleftrightarrow x = y]$,
which is usable as a rewrite rule.
If $f$ operates on values of a type (predicate) $\tau$,
its injectivity restricted to values of that type can be stated as
$[\tau(x) \wedge \tau(y) \Longrightarrow
 (f(x) = f(y) \Longleftrightarrow x = y)]$;
if $f$ implicitly or explicitly fixes values outside $\tau$
via a fixer $\phi$ for $\tau$,
injectivity can be equivalently stated as
$[f(x) = f(y) \Longleftrightarrow \phi(x) = \phi(y)]$,%
\footnote{This is obtained by
replacing $x$ and $y$ with $\phi(x)$ and $\phi(y$) in
$[\tau(x) \wedge \tau(y) \Longrightarrow
 (f(x) = f(y) \Longleftrightarrow x = y)]$,
and using the fact that
$\forall z. \tau(\phi(z))$ (i.e.\ $\phi$ is a fixer for $\tau$) and
$\forall z. f(\phi(z)) = f(z)$ (i.e.\ $f$ fixes its argument).
Conversely, assuming $\tau(x)$ and $\tau(y)$,
$[f(x) = f(y) \Longleftrightarrow \phi(x) = \phi(y)]$
reduces to $[f(x) = f(y) \Longleftrightarrow x = y]$
because $\forall z. \tau(z) \Longrightarrow \phi(z) = z$
(i.e.\ $\phi$ is identity over $\tau$).}
which is generally preferable as a rewrite rule because it has no hypotheses.
This is the formulation in \figref{fig:rlp-decodable-acl2},
applied to the \acl{encoding} results and
with the necessary hypotheses that the \acl{error?} results are \acl{nil};
the results are obtained via \acl{mv-nth}
\citeman{http://www.cs.utexas.edu/users/moore/acl2/manuals/current/manual/?topic=ACL2____MV-NTH}{mv-nth}.

First, a theorem \acl{rlp-encode-bytes-injective},
not shown here but analogous to \acl{rlp-encode\-tree-injective}
(with \acl{rlp-encode-tree} and \acl{rlp-tree-fix}
replaced with \acl{rlp-encode-bytes} and \acl{byte-list-fix}),
is proved by first proving a variant lemma (not shown)
with \acl{byte-listp} hypotheses on \acl{x} and \acl{y}
and without \acl{byte-list-fix}
(i.e.\ the form
$[\tau(x) \wedge \tau(y) \Longrightarrow
 (f(x) = f(y) \Longleftrightarrow x = y)]$
above).
The lemma is proved automatically,
once the definition of \acl{rlp-encode-bytes} is enabled via a hint,
which leads ACL2 to consider nine cases---%
three for \acl{x} and three for \acl{y},
corresponding to the three branches
in the definition of \acl{rlp-encode-bytes};
see the documentation
\citeman{http://www.cs.utexas.edu/users/moore/acl2/manuals/current/manual/?topic=ETHEREUM\_\_\_\_RLP}{rlp}
for details.
Then the theorem is easily proved from the lemma
via a hint (not shown)
to use the instance of the lemma where \acl{x} and \acl{y}
are replaced with \acl{(byte-list-fix x)} and \acl{(byte-list-fix y)}.
Attempting to prove the theorem directly,
with just the hint to enable the definition of \acl{rlp-encode-bytes},
fails.

Then, the theorem \acl{rlp-encode-tree-injective} is also proved
by first proving a variant lemma
analogous to the one for \acl{rlp-encode-bytes-injective} described above
(i.e.\ the form
$[\tau(x) \wedge \tau(y) \Longrightarrow
 (f(x) = f(y) \Longleftrightarrow x = y)]$
above).
Since \acl{rlp-encode-tree}
is mutually recursive with \acl{rlp-encode\-tree-list},
this lemma is proved by induction
along with a similar lemma about the injectivity of \acl{rlp-encode-tree-list}
(not shown).
The induction schemas that ACL2 automatically generates
for trees and for the encoding functions,
which operate on single variables,
do not work for these lemmas:
a new induction schema (not shown) is provided
that operates on two variables simultaneously,
such as \acl{x} and \acl{y} in the lemmas;
this new induction schema is more general than its use in this proof.
Some of the base and step cases of the induction are proved automatically,
while others make use of some fairly specific lemmas (not shown);
see the documentation
\citeman{http://www.cs.utexas.edu/users/moore/acl2/manuals/current/manual/?topic=ETHEREUM\_\_\_\_RLP}{rlp}
for details.
The base case, in which \acl{x} or \acl{y} is a leaf tree,
makes use of \acl{rlp-encode-bytes-injective},
the injectivity theorem for \acl{rlp-encode-bytes}.

The injectivity of the RLP encoding functions
could be alternatively proved by
defining RLP decoding functions,
proving that the latter are left inverses of the former,
and using the inverse property to prove injectivity.%
\footnote{If $f$ has a left inverse $g$,
then given $f(x) = f(y)$,
and applying $g$ to both sides to obtain $g(f(x)) = g(f(y))$,
the left inverse property yields $g(f(x)) = x = y = g(f(y))$.
Thus $f$ is injective.}
In contrast, the injectivity proofs explained above,
which are not particularly difficult,
are solely in terms of the encoding functions,
which is more elegant and abstract in my opinion.

The prefix-unambiguity theorem in \figref{fig:rlp-decodable-acl2}
says that if a valid encoding is a prefix of another valid encoding,
then the two encodings are equal;
similarly to the injectivity theorem,
this theorem also states the obvious converse implication,
making the theorem a more useful rewrite rule.
The library function \acl{prefixp} tests whether the first argument
is a (not necessarily strict) prefix of the second argument.
Thus, the theorem prohibits a valid encoding
from being a strict prefix of another valid encoding.

The theorem \acl{rlp-encode-tree-unamb-prefix}
is proved via a case split, specified as a hint (not shown),
on whether the lengths of the encodings are equal or not.
If the lengths are equal,
the encodings must be equal since one is a prefix of the other;
this is proved automatically via library theorems about \acl{prefixp}.
If the lengths are not equal,
a more general theorem (not shown) is used
to show that the lengths must be actually equal,
proving this case by contradiction.
That more general theorem says that
the length of an encoding is determined by the first few bytes of the encoding,
because encodings start with length information.
Additional hints (not shown) guide ACL2 to
recognizing that those first few bytes must be the same for the two encodings
if one is a prefix of the other.


\section{RLP Decoding}


\subsection{Declarative Specification}
\label{sec:declarative}

The RLP decoding of trees from their encodings
is formalized in \figref{fig:rlp-decode-decl-acl2}.

\begin{figure}
\centering
\begin{minipage}{0.64\textwidth}
\begin{bacl}
(define rlp-decode-tree ((encoding byte-listp))
  :returns (mv (error? booleanp) (tree rlp-treep))
  (b* ((encoding (byte-list-fix encoding)))
    (if (rlp-tree-encoding-p encoding)
        (mv nil (rlp-tree-encoding-witness encoding))
      (mv t (rlp-tree-leaf nil))))) ; 2nd result irrelevant
\end{bacl}
\end{minipage}
\caption{\label{fig:rlp-decode-decl-acl2}
Declarative definition of RLP decoding in ACL2.}
\end{figure}

The function \acl{rlp-decode-tree}
takes as input a list of bytes that is the purported encoding
and returns as output a pair consisting of
a boolean error flag and a tree.
The error flag is \acl{nil} exactly when
the input is a valid encoding of a tree:
in this case, that tree is returned,
via the witness function \acl{rlp-tree-encoding-witness}
associated to \acl{rlp-tree-encoding-p}
in \figref{fig:valid-rlp-encodings-acl2}
(more details on this below).
Otherwise, the error flag is \acl{t}
and the second component of the output is irrelevant.

The existentially quantified function \acl{rlp-tree-encoding-p}
is defined in terms of the witness function \acl{rlp-tree-encoding-witness}
(under the hood 
\citeman{http://www.cs.utexas.edu/users/moore/acl2/manuals/current/manual/?topic=ACL2____DEFUN-SK}{defun-sk};
see \acl{:skolem-name} in \figref{fig:valid-rlp-encodings-acl2}),
which is axiomatized, via the matrix of the quantification,
to be a right inverse of the encoding function \acl{rlp-encode-tree}:
if \acl{encoding} is a valid tree encoding,
then \acl{(rlp-tree-encoding-witness encoding)} returns
\acl{tree} of type \acl{rlp-treep} such that
\acl{(rlp-encode-tree tree)} returns \acl{(mv nil encoding)},
i.e.\ no error and the original encoding.
This readily implies that \acl{rlp-decode-tree} is
a right inverse of \acl{rlp-encode-tree},
as stated by the theorem \acl{rlp-encode-tree-of-rlp-decode-tree}
in \figref{fig:rlp-tree-encode-decode-inverses-acl2}:
if \acl{encoding} is a valid tree encoding (modulo \acl{byte-list-fix}),
then \acl{rlp-decode-tree} succeeds and returns \acl{tree},
and \acl{rlp-encode-tree} succeeds on \acl{tree}
and returns the original encoding.
This is proved automatically,
using easily proved theorems (not shown) about
\acl{rlp-tree-encoding-witness}.

\begin{figure}
\centering
\begin{minipage}{0.7\textwidth}
\begin{bacl}
(defthm rlp-encode-tree-of-rlp-decode-tree
  (implies (rlp-tree-encoding-p encoding)
           (b* (((mv d-error? tree) (rlp-decode-tree encoding))
                ((mv e-error? encoding1) (rlp-encode-tree tree)))
             (and (not d-error?)
                  (not e-error?)
                  (equal encoding1 (byte-list-fix encoding))))))

(defthm rlp-decode-tree-of-rlp-encode-tree
  (b* (((mv e-error? encoding) (rlp-encode-tree tree))
       ((mv d-error? tree1) (rlp-decode-tree encoding)))
    (implies (not e-error?)
             (and (not d-error?)
                  (equal tree1 (rlp-tree-fix tree))))))
\end{bacl}
\end{minipage}
\caption{\label{fig:rlp-tree-encode-decode-inverses-acl2}
ACL2 theorems asserting that RLP encoding and decoding are mutual inverses.}
\end{figure}

Since, as discussed in \secref{sec:decodability},
\acl{rlp-encode-tree} is injective,
\acl{rlp-decode-tree} is also
a left inverse of \acl{rlp-encode-tree},%
\footnote{In general,
if a function $f$ is injective and has a right inverse $g$,
then $g$ is also a left inverse of $f$:
given the right inverse property $f(g(x)) = x$,
replacing $x$ with $f(y)$ yields $f(g(f(y))) = f(y)$,
which the injectivity of $f$ reduces to $g(f(y)) = y$,
i.e.\ the left inverse property.}
as stated by the theorem \acl{rlp-decode-tree-of-rlp-encode-tree}
in \figref{fig:rlp-tree-encode-decode-inverses-acl2}:
if \acl{tree} is (modulo \acl{rlp-tree-fix}) a tree,
and \acl{rlp-encode-tree} succeeds and returns \acl{encoding},
then \acl{rlp-decode-tree} succeeds on \acl{encoding}
and returns the original tree.
This is proved via a few hints (not shown)
to instantiate
(replacing \acl{encoding} with \acl{(mv-nth 1 (rlp-encode-tree tree))})
and use
the right inverse theorem
\acl{rlp-encode-tree-of-rlp-decode-tree},
while the injectivity theorem in \figref{fig:rlp-decodable-acl2}
is automatically used as a rewrite rule.

\figref{fig:rlp-decode-decl-acl2} defines the RLP decoding function
as inverse of the RLP encoding function.
This is a declarative, non-executable definition.
YP does not explicitly define any RLP decoding function,
but, clearly, it implicitly defines it as inverse of the encoding function:
this implicit definition is formalized in \figref{fig:rlp-decode-decl-acl2}.
WK provides Python reference code for RLP decoding,
but the definition in \figref{fig:rlp-decode-decl-acl2}
is more abstract and manifestly correct.


\subsection{Executable Implementation}
\label{sec:implementation}

The function \acl{rlp-decode-tree} in \figref{fig:rlp-decode-decl-acl2}
can be regarded as a high-level specification of RLP decoding,
which can be implemented by an equivalent executable function.

The executable decoding function is defined
in terms of the executable RLP parser shown in
\figref{fig:rlp-parse-tree-acl2}.
This function is mutually recursive with the one in
\figref{fig:rlp-parse-tree-list-acl2}.
Similarly to \figref{fig:rlp-encode-acl2},
these two functions are surrounded by \acl{defines} (not shown).

\begin{figure}
\centering
\begin{minipage}{0.87\textwidth}
\begin{bacl}
(define rlp-parse-tree ((encoding byte-listp))
  :returns (mv (error? maybe-rlp-error-p) (tree rlp-treep) (rest byte-listp))
  (b* ((encoding (byte-list-fix encoding))
       ((when (endp encoding)) ...) ; error
       ((cons first encoding) encoding)
       ((when (< first 128)) (mv nil (rlp-tree-leaf (list first)) encoding))
       ((when (<= first 183))
        (b* ((len (- first 128))
             ((when (< (len encoding) len)) ...) ; error
             (bytes (take len encoding))
             ((when (and (= len 1) (< (car bytes) 128))) ...)) ; error
          (mv nil (rlp-tree-leaf bytes) (nthcdr len encoding))))
       ((when (< first 192))
        (b* ((lenlen (- first 183))
             ((when (< (len encoding) lenlen)) ...) ; error
             (len-bytes (take lenlen encoding))
             ((unless (equal (trim-bendian* len-bytes) len-bytes)) ...) ; error
             (encoding (nthcdr lenlen encoding))
             (len (bebytes=>nat len-bytes))
             ((when (<= len 55)) ...) ; error
             ((when (< (len encoding) len)) ...) ; error
             (bytes (take len encoding))
             (encoding (nthcdr len encoding)))
          (mv nil (rlp-tree-leaf bytes) encoding)))
       ((when (<= first 247))
        (b* ((len (- first 192))
             ((when (< (len encoding) len)) ...) ; error
             (subencoding (take len encoding))
             (encoding (nthcdr len encoding))
             ((mv error? subtrees) (rlp-parse-tree-list subencoding))
             ((when error?) ...)) ; error
          (mv nil (rlp-tree-branch subtrees) encoding)))
       (lenlen (- first 247))
       ((when (< (len encoding) lenlen)) ...) ; error
       (len-bytes (take lenlen encoding))
       ((unless (equal (trim-bendian* len-bytes) len-bytes)) ...) ; error
       (encoding (nthcdr lenlen encoding))
       (len (bebytes=>nat len-bytes))
       ((when (<= len 55)) ...) ; error
       ((when (< (len encoding) len)) ...) ; error
       (subencoding (take len encoding))
       (encoding (nthcdr len encoding))
       ((mv error? subtrees) (rlp-parse-tree-list subencoding))
       ((when error?) ...)) ; error
    (mv nil (rlp-tree-branch subtrees) encoding)))
\end{bacl}
\end{minipage}
\caption{\label{fig:rlp-parse-tree-acl2}
Executable parser of RLP encodings in ACL2 (Part 1).}
\end{figure}

\begin{figure}
\centering
\begin{minipage}{0.76\textwidth}
\begin{bacl}
(define rlp-parse-tree-list ((encoding byte-listp))
  :returns (mv (error? maybe-rlp-error-p) (trees rlp-tree-listp))
  (b* (((when (endp encoding)) (mv nil nil))
       ((mv error? tree encoding1) (rlp-parse-tree encoding))
       ((when error?) ...) ; error
       ((unless (mbt (< (len encoding1) (len encoding)))) ...) ; error
       ((mv error? trees) (rlp-parse-tree-list encoding1))
       ((when error?) ...)) ; error
    (mv nil (cons tree trees))))
\end{bacl}
\end{minipage}
\caption{\label{fig:rlp-parse-tree-list-acl2}
Executable parser of RLP encodings in ACL2 (Part 2).}
\end{figure}

The function \acl{rlp-parse-tree}
takes as input a list of bytes
and returns as output a triple consisting of
an error indication (\acl{nil} if parsing is successful),
a decoded tree (an irrelevant one if parsing fails),
and the remaining bytes after the parsed encoding
(\acl{nil} if parsing fails%
\footnote{Returning \acl{nil} as third result when parsing fail
may seem an odd choice,
compared to returning the remaining bytes
whose parsing caused the error,
which could convey information about the error.
However, this (and more) information is already included in the error result,
whose recognizer \acl{rlp-error-p} is mentioned a few sentences later
(but not explained in detail).}).
The parser stops as soon as a tree is successfully decoded
(because of theorem \acl{rlp-encode-tree-unamb-prefix}
in \figref{fig:rlp-decodable-acl2},
there cannot be a longer encoding to parse),
returning the remaining bytes for further parsing:
thus, as lists of trees are recursively parsed,
the input bytes are threaded through, and consumed chunk-wise.
The function \acl{rlp-parse-tree-list}
takes as input a list of bytes
and returns as output a pair consisting of
an error indication (similarly to \acl{rlp-parse-tree})
and a list of decoded trees;
it returns no remaining bytes
because it is always called (by \acl{rlp-parse-tree})
on a sublist of the input of known length
that must exactly encode zero or more trees.
The predicate \acl{maybe-rlp-error-p}
recognizes \acl{nil} (for no error)
and error values (recognized by \acl{rlp-error-p}, not shown)
that convey information about different possible parsing errors;
the exact errors are not shown in \figref{fig:rlp-parse-tree-acl2},
replaced with ellipses.

The parser operates as follows
(cf.\ the encoding functions in \figref{fig:rlp-encode-acl2}):
\begin{enumerate}[nosep]
\item
It fixes the \acl{encoding} input to be a list of bytes
(a no-op under the guard).
\item
If the input list of bytes is empty, an error is returned:
RLP encodings are never empty.
\item
Otherwise,
the \acl{encoding} input is matched to the \acl{(cons first encoding)} pattern,
binding the variable \acl{first} to the first byte of the encoding
and the (new) variable \acl{encoding} to the rest of the encoding.
This way, the first byte can be examined to determine what to do next.
\item
If the first byte is below 128,
it encodes the singleton list of itself,
and so the byte is returned as a leaf tree.
\item
If the first byte is in the range 128--183,
the encoding must be of a byte array with length below 56.
The length is calculated:
if not enough bytes occur after the first, an error is returned;
otherwise, the bytes are returned as a leaf tree.
An error is also returned if there is just one byte below 128;
see \secref{sec:quasi-encodings}.
\item
If the first byte of the encoding is in the range 184--191,
the encoding must be of a byte array
whose length \acl{len} is encoded by
the next \acl{lenlen} bytes after the first one,
as a base-256 big-endian no-leading-zeros list.
After obtaining \acl{lenlen} from the first byte,
an error is returned if there are not enough bytes to encode the length.
An error is also returned if the encoded length has leading zeros
(the library function \acl{trim-bendian*}
removes all the leading zeros from a list of big-endian digits);
see \secref{sec:quasi-encodings}.
Otherwise, the library function \acl{bebytes=>nat},
inverse of \acl{nat=>bebytes*} in \figref{fig:rlp-encode-acl2},
is used to calculate the length \acl{len} of the encoded byte array.
An error is returned if this length is below 56;
see \secref{sec:quasi-encodings}.
If there are enough bytes after the first and the next \acl{lenlen} bytes,
they are returned as a leaf tree.
\item
If the first byte of the encoding is in the range 192--247,
the encoding must be of a branching tree
whose subtrees have a total encoded length below 56.
That exact number of bytes is passed to \acl{rlp-parse-tree-list}
(which is described in more detail below),
which returns an error indication (\acl{nil} if no error)
and the list of decoded trees,
which are wrapped into a branching tree and returned.
As above, an error is returned if there are not enough bytes in the input.
Any error from \acl{rlp-parse-tree-list} is returned by \acl{rlp-parse-tree}.
\item
If the first byte of the encoding is in the range 248--255,
the encoding must be of a branching tree
whose length \acl{len} is encoded by
the next \acl{lenlen} bytes after the first one,
as a base-256 big-endian no-leading-zeros list.
The processing of \acl{lenlen} and \acl{len}
is analogous to the case in which the first byte is in the range 184--191
(explained above),
including returning errors if there are leading zeros in the length
or if \acl{len} is below 56 (see \secref{sec:quasi-encodings}).
Similarly to the case in which the first byte is in the range 192--247,
\acl{rlp-parse-tree-list} is called with the exact number of bytes to parse,
any error from that call is propagated,
and ultimately (if no errors occur) the decoded branching tree is returned.
\end{enumerate}

The function \acl{rlp-parse-tree-list} in \figref{fig:rlp-parse-tree-list-acl2}
takes as input a list of bytes purported to encode zero or more trees
and returns as output a pair consisting of
an error indication (\acl{nil} if no error)
and a list of decoded RLP trees.
Unlike \acl{rlp-parse-tree}, it does not return any remaining input bytes,
as explained earlier.

The body of \acl{rlp-parse-tree-list} decodes trees
while there are input bytes available,
stopping if an error occurs and propagating that error.
After each tree is decoded, the remaining bytes are recursively parsed.
The \acl{mbt} expression serves to prove termination, as discussed below.

The termination of the mutually recursive functions
\acl{rlp-parse-tree} and \acl{rlp-parse-tree-list}
is proved automatically once a measure (not shown) is provided.
The measure is lexicographic:
it consists of the length of the input list of bytes,
followed by a linear ordering of the two functions
defined by \acl{rlp-parse-tree} being smaller than \acl{rlp-parse-tree-list}.
When \acl{rlp-parse-tree} calls \acl{rlp-parse-tree-list},
the first component of the measure decreases.
When \acl{rlp-parse-tree-list} calls \acl{rlp-parse-tree},
the first component is unchanged but the second decreases.
When \acl{rlp-parse\-tree-list} calls itself,
the first component also decreases,
because RLP encodings are never empty,
and therefore the preceding call of \acl{rlp-parse-tree}
must have consumed some input bytes.
The latter is a property of \acl{rlp-parse-tree},
but in order to prove this property,
the function must be first accepted by ACL2,
requiring its termination to be proved.
This circularity is broken via the run-time test
\acl{(< (len encoding1) (len encoding))},
which gets the function definition accepted by ACL2.
The \acl{mbt} around the test avoids executing the test at run time
\citeman{http://www.cs.utexas.edu/users/moore/acl2/manuals/current/manual/?topic=ACL2____MBT}{mbt}.

The guard verification proofs,
which involve the truth of the aforementioned \acl{mbt} test,
are automatic after proving a theorem saying that
the third result of \acl{rlp-parse-tree}, i.e.\ the remaining input bytes,
is strictly shorter than the input.
This theorem (not shown) is proved automatically
after supplying a hint to expand the definition of \acl{rlp-parse-tree},
which otherwise ACL2's heuristics apparently prevent from expanding.

The executable RLP decoding function for trees
is defined in \figref{fig:rlp-decode-exec-acl2};
the `\acl{x}' in its name stands for `executable',
as opposed to the declaratively defined decoding function
in \figref{fig:rlp-decode-decl-acl2}.
This decoding function takes as input a list of bytes,
purported to be a complete encoding with no extra bytes,%
\footnote{This executable decoding function is meaningful
as implementation of the inverse of the encoding function
declaratively defined in \figref{fig:rlp-decode-decl-acl2}.
It can be used when the lengths of the purported encodings are known.
When the lengths are not known, e.g.\ when decoding from byte streams,
the executable parser in \figref{fig:rlp-parse-tree-acl2} can be used instead.}
and returns as output a pair consisting of an error indication
(\acl{nil} if decoding is successful)
and a tree---whose value is irrelevant if the first result is not \acl{nil}.
A tree is decoded by calling the parser
and ensuring that there are no remaining bytes.

\begin{figure}
\centering
\begin{minipage}{0.65\textwidth}
\begin{bacl}
(define rlp-decodex-tree ((encoding byte-listp))
  :returns (mv (error? maybe-rlp-error-p) (tree rlp-treep))
  (b* (((mv error? tree rest) (rlp-parse-tree encoding))
       ((when error?) ...) ; error
       ((when (consp rest)) ...)) ; error
    (mv nil tree)))
\end{bacl}
\end{minipage}
\caption{\label{fig:rlp-decode-exec-acl2}
Executable definition of RLP decoding in ACL2.}
\end{figure}


\subsubsection{Rejection of Invalid Quasi-Encodings}
\label{sec:quasi-encodings}

As mentioned above,
the parser rejects ``quasi-encodings'' of the following forms:%
\begin{itemize}[nosep]
\item
$\rlpstring{129,x}$
with $x<128$:
this ``could'' encode
(a leaf tree consisting of) a singleton byte array $\rlpstring{x}$,
but RLP prescribes the encoding $\rlpstring{x}$ in this case.
\item
$\rlpstring{183+\lenlen,\len_1,\ldots,\len_\lenlen,x,\ldots}$
with $1 \leq \lenlen \leq 8$ and $\len_1 = 0$:
this ``could'' encode
(a leaf tree consisting of) a byte array $\rlpstring{x,\ldots}$ of length
$\len = \sum_{1 \leq i \leq \lenlen} \len_i \times 256^{\lenlen - i}$,
but RLP prescribes the absence of leading zeros
in the base-256 big-endian representation of $\len$.
\item
$\rlpstring{184,\len,x,\ldots}$
with $0<\len<56$:
this ``could'' encode
(a leaf tree consisting of) a byte array $\rlpstring{x,\ldots}$
of length $\len$,
but RLP prescribes the encoding $\rlpstring{128+\len,x,\ldots}$ in this case.
\item
$\rlpstring{247+\lenlen,\len_1,\ldots,\len_\lenlen,x,\ldots}$
with $1 \leq \lenlen \leq 8$ and $\len_1 = 0$:
this ``could'' encode
a branching tree whose concatenated encoded subtrees have length
$\len = \sum_{1 \leq i \leq \lenlen} \len_i \times 256^{\lenlen - i}$,
but RLP prescribes the absence of leading zeros
in the base-256 big-endian representation of $\len$.
\item
$\rlpstring{248,\len,x,\ldots}$
with $0<\len<56$:
this ``could'' encode
a branching tree whose concatenated encoded subtrees have length $\len$,
but RLP prescribes the encoding $\rlpstring{192+\len,x,\ldots}$ in this case.
\end{itemize}
These are not valid encodings because
they do not satisfy the predicate in \figref{fig:valid-rlp-encodings-acl2}.
While they ``could'' be encodings in the sense mentioned above,
they are ``non-optimal'': shorter valid encodings exist.

When implementing RLP decoding,
the rejection of these quasi-encodings may be easily overlooked.
My initial writing of the parser failed to reject these quasi-encodings;
I discovered the problem when failing to prove
the right inverse property in \secref{sec:verification}.
The Python reference code in WK used to accept these quasi-encodings,
and some existing RLP implementations
used to accept or still accept them as well (see \secref{sec:related}).

While it may seem benign to accept these quasi-encodings,
they are not in the image of the RLP encoding function.
Also see \url{https://gitter.im/ethereum/research/archives/2016/07/30};
search for `consensus rule'.
A possible problematic scenario
is a database that uses RLP encodings as keys:
if different encodings for the same item were accepted,
the item could be stored multiple times in the database.


\subsection{Verification of Correctness}
\label{sec:verification}

The first step toward proving that
the executable definition in \figref{fig:rlp-decode-exec-acl2}
is equivalent to
the declarative definition in \figref{fig:rlp-decode-decl-acl2}
is to prove that the tree parsing and encoding functions are mutual inverses,
which is expressed by
the theorems in \figref{fig:rlp-tree-encode-parse-inverses-acl2}.

\begin{figure}
\centering
\begin{minipage}{0.65\textwidth}
\begin{bacl}
(defthm rlp-parse-tree-of-rlp-encode-tree
  (b* (((mv e-error? encoding) (rlp-encode-tree tree))
       ((mv d-error? tree1 rest) (rlp-parse-tree encoding)))
    (implies (not e-error?)
             (and (not d-error?)
                  (not (consp rest))
                  (equal tree1 (rlp-tree-fix tree))))))

(defthm rlp-encode-tree-of-rlp-parse-tree
  (b* (((mv d-error? tree rest) (rlp-parse-tree encoding))
       ((mv e-error? encoding1) (rlp-encode-tree tree)))
    (implies (not d-error?)
             (and (not e-error?)
                  (equal (append encoding1 rest)
                         (byte-list-fix encoding))))))
\end{bacl}
\end{minipage}
\caption{\label{fig:rlp-tree-encode-parse-inverses-acl2}
ACL2 theorems asserting that RLP encoding and parsing are mutual inverses.}
\end{figure}

The first theorem in \figref{fig:rlp-tree-encode-parse-inverses-acl2}
says that \acl{rlp-parse-tree} is a left inverse of \acl{rlp-encode-tree}
over the encodable trees,
i.e.\ the parser recognizes and reconstructs
\emph{all} valid encodings of trees.
More in detail, it says that
if \acl{rlp-encode-tree} succeeds,
then \acl{rlp-parse-tree} succeeds on the resulting encoding,
returning the original tree (modulo fixing)
and no remaining bytes (i.e.\ it consumes the whole encoding).
Since \acl{rlp-parse-tree} is mutually recursive with \acl{rlp-parse-tree-list},
and \acl{rlp-encode-tree} is mutually recursive with \acl{rlp-encode-tree-list},
that theorem is proved at the same time as
another theorem (not shown)
that says that \acl{rlp-parse-tree-list} is
a left inverse of \acl{rlp-encode-tree-list}.
These two theorems are proved by induction on
\acl{rlp-encode-tree} and \acl{rlp-encode-tree-list},
via the \acl{make-flag} macro
\citeman{http://www.cs.utexas.edu/users/moore/acl2/manuals/current/manual/?topic=ACL2\_\_\_\_MAKE-FLAG}{make-flag}.
The proof is automatic
once the definitions of these and other functions are enabled,
and some hints (not shown) are provided
to force the expansion of some calls of \acl{rlp-parse-tree};
see the documentation
\citeman{http://www.cs.utexas.edu/users/moore/acl2/manuals/current/manual/?topic=ETHEREUM\_\_\_\_RLP}{rlp}
for details.
The proof also makes use of a previously proved more general rewrite rule
(not shown)
saying that if \acl{rlp-parse-tree} succeeds on \acl{encoding},
returning a tree and some remaining bytes \acl{rest},
it also succeeds on an extended input \acl{(append encoding more-bytes)},
returning the same tree and \acl{(append rest more-bytes)} as remaining bytes.

The second theorem in \figref{fig:rlp-tree-encode-parse-inverses-acl2}
says that \acl{rlp-parse-tree} is a right inverse of \acl{rlp-encode-tree}
over the valid tree encodings,
i.e.\ the parser recognizes and reconstructs
\emph{only} valid encodings of trees:
if it accepted an invalid encoding and returned a tree,
\acl{rlp-encode-tree} would have to map that tree back to the encoding,
which would be therefore valid, contradicting the hypothesis that it is invalid.
More in detail, the theorem says that
if \acl{rlp-parse-tree} succeeds,
then \acl{rlp-encode-tree} succeeds on the resulting tree,
returning the prefix of the original encoding (modulo fixing)
that omits the remaining bytes returned by \acl{rlp-parse-tree}.
Analogously to the left inverse theorem described above,
this theorem is proved at the same time as
another theorem (not shown) about
\acl{rlp-parse-tree-list} and \acl{rlp-encode-tree-list},
via \acl{make-flag};
the proof,
by induction on \acl{rlp-parse-tree}
and \acl{rlp-parse\-tree-list}%
\footnote{In ACL2, as in NQHTM \cite[Chapt.~15]{acl-book},
induction is applicable
when the arguments that decrease in the recursion are variables.
Thus, while the left inverse theorems are proved
by induction on \acl{rlp-encode-tree} and \acl{rlp-encode-tree-list},
the right inverse theorems are proved
by induction on \acl{rlp-parse-tree}
and \acl{rlp-parse-tree-list}.},
is automatic once the definitions of these and other functions are enabled.

If the parser accepted the quasi-encodings
discussed in \secref{sec:quasi-encodings},
the left inverse theorem would still hold,
but the right inverse theorem would not.

The fact that \acl{rlp-decodex-tree} in \figref{fig:rlp-decode-exec-acl2}
is both a left and a right inverse of \acl{rlp-encode-tree}
easily follows from
the theorems in \figref{fig:rlp-tree-encode-parse-inverses-acl2}.
This fact is asserted by two theorems (not shown)
that are written as in \figref{fig:rlp-tree-encode-decode-inverses-acl2}
but where \acl{rlp-decode-tree} is replaced with \acl{rlp-decodex-tree}.
The left inverse theorem is proved automatically
once the definition of \acl{rlp-decodex-tree} is enabled;
the first theorem in \figref{fig:rlp-tree-encode-parse-inverses-acl2}
applies as a rewrite rule.
The right inverse theorem,
besides enabling the definition of \acl{rlp-decodex-tree},
requires a couple of hints (not shown)
to use the second theorem in \figref{fig:rlp-tree-encode-parse-inverses-acl2},
since the \acl{append} in it
does not make it readily applicable as a rewrite rule in this case.

The equivalence of \acl{rlp-decodex-tree} and \acl{rlp-decode-tree}
is stated by the theorem in \figref{fig:rlp-tree-decodex-correct-acl2}.
Since \acl{rlp-decode-tree} returns a boolean error result
while \acl{rlp-decodex-tree} returns a richer range of error results,
the first results of these two functions are only \acl{iff}-equivalent,
i.e.\ one is \acl{nil} if and only if the other one is \acl{nil};
their second results are always equal instead.

\begin{figure}
\centering
\begin{minipage}{0.6\textwidth}
\begin{bacl}
(defthm rlp-decode-tree-is-rlp-decodex-tree
  (and (iff (mv-nth 0 (rlp-decode-tree encoding))
            (mv-nth 0 (rlp-decodex-tree encoding)))
       (equal (mv-nth 1 (rlp-decode-tree encoding))
              (mv-nth 1 (rlp-decodex-tree encoding)))))
\end{bacl}
\end{minipage}
\caption{\label{fig:rlp-tree-decodex-correct-acl2}
ACL2 theorem asserting the correctness of the executable RLP decoder.}
\end{figure}

The theorem in \figref{fig:rlp-tree-decodex-correct-acl2}
is proved by cases on whether \acl{(rlp-tree-encoding-p encoding)} holds or not.
A preliminary lemma (not shown) is proved,
with a few hints and a couple of simple intermediate lemmas,
asserting the equivalence of
(i) \acl{rlp-tree-encoding-p} returning \acl{t} and
(ii) \acl{rlp-decodex-tree} returning a \acl{nil} error result.
If \acl{rlp-tree-encoding-p} holds:
\begin{itemize}[nosep]
\item
The first conjunct in \figref{fig:rlp-tree-decodex-correct-acl2}
is proved via a few hints,
using the aforementioned lemma and the definition of \acl{rlp-decode}.
\item
The second conjunct in \figref{fig:rlp-tree-decodex-correct-acl2}
is proved via a few hints,
from \acl{rlp-encode-tree-injective}
and the right inverse properties of
\acl{rlp-decode-tree} and \acl{rlp-decodex-tree}.%
\footnote{In general,
if an injective function $f$ has right inverses $g$ and $h$, then $g = h$:
from the right inverse properties $f(g(x)) = x = f(h(x))$,
injectivity gives $g(x) = h(x)$.}
\end{itemize}
If \acl{rlp-tree-encoding-p} does not hold,
both conjuncts are proved via a few hints,
using the aforementioned preliminary lemma
and the definitions of \acl{rlp-decode-tree} and \acl{rlp-decodex-tree}.


\section{Related Work and Impact}
\label{sec:related}

RLP is formally defined in YP.%
\footnote{Using ``pencil-and-paper'' mathematical notation,
not in a theorem prover or similar tool.}
Based on the ACL2 development,
I contributed some improvements to that definition
(see Pull Requests
\href{https://github.com/ethereum/yellowpaper/pull/700}{700},
\href{https://github.com/ethereum/yellowpaper/pull/736}{736},
\href{https://github.com/ethereum/yellowpaper/pull/739}{739},
\href{https://github.com/ethereum/yellowpaper/pull/742}{742},
\href{https://github.com/ethereum/yellowpaper/pull/745}{745}, and
\href{https://github.com/ethereum/yellowpaper/pull/746}{746}
at \url{https://github.com/ethereum/yellowpaper})
and helped close some outstanding items
(see Pull Request \href{https://github.com/ethereum/yellowpaper/pull/648}{648}
and Issue \href{https://github.com/ethereum/yellowpaper/issues/116}{116}
at \url{https://github.com/ethereum/yellowpaper}).

RLP is informally defined in WK.
Previously, the Python reference code for RLP decoding in WK
accepted the quasi-encodings described in \secref{sec:quasi-encodings}.
Based on the ACL2 development,
I contributed a fix to reject those quasi-encodings
(see Issue \href{https://github.com/ethereum/wiki/issues/668}{688}
at \url{https://github.com/ethereum/wiki}).

The KEVM \cite{kevm-techrep} \cite{jello-paper} includes
executable specifications of RLP encoding and decoding.%
\footnote{The KEVM is an evolving artifact.
The following assertions are current at the time of this writing.}
The decoding specification covers all encodings,
while the encoding specification only covers byte arrays
and some data types that are encoded like byte arrays.
There are no theorems stating that
the specified encoding and decoding are mutual inverses.
The decoding specification appears to accept
some of the quasi-encodings described in \secref{sec:quasi-encodings}
(see Issue
\href{https://github.com/kframework/evm-semantics/issues/413}{413}
at \url{https://github.com/kframework/evm-semantics}).

The Lem EVM \cite{lem-evm-paper} \cite{lem-evm-code}
includes a partial (apparently in progress) Isabelle/HOL specification
of RLP encoding and decoding.%
\footnote{The Lem EVM is an evolving artifact.
The following assertions are current at the time of this writing.}
The type of RLP trees is much like \figref{fig:rlp-trees-acl2}.
There is a complete specification of RLP encoding,
but only a partial specification of RLP decoding.
There are no theorems stating that
the specified encoding and decoding are mutual inverses.

There are several implementations of RLP,
in libraries and Ethereum clients,
written in mainstream programming languages.
Some of these implementations used to accept, or still accept,
the quasi-encodings described in \secref{sec:quasi-encodings}
(e.g.\ see Issue
\href{https://github.com/ethereum/aleth/issues/1639}{1639}
at \url{https://github.com/ethereum/aleth}
and Issue
\href{https://github.com/paritytech/parity-common/issues/49}{49}
at \url{https://github.com/paritytech/parity-common}).

The Ethereum test suite \cite{eth-tests}
contains JSON-formatted tests for all Ethereum implementations,
including tests for RLP encoding and decoding.
Previously,
this test suite included a few tests for rejecting some, but not all, of
the five kinds of quasi-encodings described in \secref{sec:quasi-encodings}.
Based on the ACL2 development,
I contributed additional tests
to cover all the five kinds of quasi-encodings
(see Pull Request
\href{https://github.com/ethereum/tests/pull/612}{612}
at \url{https://github.com/ethereum/tests}).

There is extensive work on
formal verification and analysis of (Ethereum and other) smart contracts,
for which just some references are provided here
\cite{isabelle-eth}
\cite{coq-eth}
\cite{f-star-eth}
\cite{gasper}
\cite{cubicle}
\cite{kevm-techrep}
\cite{lem-evm-paper}
\cite{zeus}
\cite{oyente}
\cite{exact-worst-gas}
\cite{why3-eth-new}
\cite{maian}.
That work is complementary to the ACL2 RLP development,
whose focus is (a component of) the platform that runs smart contracts.
However, as mentioned in \secref{sec:prob-contrib},
this platform-focused work can support smart contract verification,
by providing the underlying formal semantics of smart contracts.


\section{Future Work}

Alternative, perhaps more efficient, implementations of
the RLP parser and decoder in \secref{sec:implementation}
could be written and verified,
comparing their proof efforts and techniques with \secref{sec:verification}.

It could be investigated how to derive, via stepwise refinement,
verified and efficient RLP parser and decoder implementations
from the declarative specification in \secref{sec:declarative},
using the APT (Automated Program Transformations) toolkit
\citeman{http://www.cs.utexas.edu/users/moore/acl2/manuals/current/manual/?topic=APT____APT}{APT}.
This may require the development of
additional, but more generally applicable, APT transformation tools.

The RLP parser and decoder in \secref{sec:implementation}
are given the complete purported encodings, or more, as input.
However, practical implementations may read the input bytes as needed,
e.g.\ from a socket.
It could be investigated how to extend
the specification, implementation, and proof
to accommodate this approach.
In this case, additional customizable length checks should be probably added,
to thwart denial-of-service-style attacks consisting in
supplying the first few bytes of very long encodings,
e.g.\ close to the $2^{64}$ limits.

The RLP parser and decoder in \secref{sec:implementation}
process encodings completely, i.e.\ to the full depth of the RLP trees.
However, practical implementations may process encodings
up to a specified tree depth initially, then to lower depths as needed.
It could be investigated how to extend
the specification, implementation, and proof
to accommodate this approach.
These new parser and decoder will need to accept invalid encodings
that are however valid up to the specified depths,
requiring a corresponding weakening of their specification.

While RLP is an important component of Ethereum,
clearly a lot more work remains
to extend the development described in this paper to a complete Ethereum client.
This remaining work is in progress
\citecode{https://github.com/acl2/acl2/tree/master/books/kestrel/ethereum}{books/kestrel/ethereum}
\citeman{http://www.cs.utexas.edu/users/moore/acl2/manuals/current/manual/?topic=ETHEREUM\_\_\_\_ETHEREUM}{ethereum}.


\section*{Acknowledgments}

\begin{itemize}[nosep]
\item
The Ethereum Foundation for supporting this work.
\item
Eric McCarthy for useful discussions on RLP
and for valuable comments on the initial draft of this paper,
including suggesting the RLP database scenario
mentioned at the end of \secref{sec:quasi-encodings}.
\item
The anonymous reviewers for useful suggestions.
\end{itemize}


\bibliographystyle{eptcs}
\bibliography{paper}


\end{document}